\documentstyle{ptptex}

\def\mat{\mib}
\newtheorem{lemma}{Lemma}      

\title{
No-Interaction Theorem \\
without Hamiltonian and Lagrangian Formalism
}
\subtitle{
Invariant Momentum on Null Cones
}    

\author{
Gen {\sc Yoneda}\footnote{E-mail: yoneda@mn.waseda.ac.jp} 
and Takesi {\sc Suzuki}
}

\inst{
Department of Mathematical Science, 
Waseda University, Tokyo 169-8555, Japan 
}

\recdate{January 27, 2000 (hep-th/0008163)
}

\abst{
In a previous paper
[G.Yoneda,
\JL{Proc.~R.~Soc.~London,A445,1994,221}.],
we proved the no-interaction theorem for four particles
with the assumption that the (linear and angular) 
momentum on space-like planes is invariant.
In this paper, we assume that 
the momentum on null cones is invariant
and prove that there is no interaction for four particles.
}

\begin{document}

\maketitle

\section{Introduction}
The no-interaction theorem \cite{Currie} states that 
there is no interaction in a relativistically invariant system 
of particles in the framework of Hamiltonian formalism.
Proofs of the no-interaction theorem have been given by many authors
\cite{Cannon,Leutwyler,Balachandran,Marmo}
in the frameworks of Hamiltonian and Lagrangian formalism.
These proofs are based on the assumption that 
there is a Hamiltonian or Lagrangian
describing the interaction between particles.

In a previous paper,\cite{Yoneda94}
we proved this theorem 
without the use of Hamiltonian or Lagrangian formalism.
The assumption in the theorem, relativistic invariance, 
is that 
the sums of the linear and angular momenta of 
individual particles on a space-like plane 
are independent of the choice of the plane.
Also, we proved that
the worldlines of the individual particles are straight.

In this paper, 
we use a null cone instead of a space-like plane.
Namely, we prove the no-interaction theorem with the
assumption of an invariant momentum on null cones.
This form of the no-interaction theorem is simple, 
like the previous form.

We consider the meaning of the no-interaction theorem 
to be summarized by the following.
\\
(a) 
Newtonian momentum invariance regarding only particles' momenta
is meaningless in special relativity.
\\
(b)
In general relativity, the flatness of Minkowski space
is usually accounted for by the observation that
zero curvature makes the worldlines of particles straight.
However, with the no-interaction theorem,
we can account for the flatness of Minkowski space 
by noting that
the invariance of the total momentum makes 
the worldlines of particles straight.
In other words, the no-interaction theorem implies 
the flatness of Minkowski space kinematically.
\\
Thus we believe that
obtaining no-interaction theorem in simple form 
is meaningful.

Next we comment on the difference between
the assumptions of this and previous papers.
Two assumptions are different with regard to geometrical meaning.
A space-like plane can be interpreted as representing simultaneity,
while a null cone does not have such a Newtonian interpretation.
A null cone can be interpreted as the set of all events that 
can be observed at some time and place.
However, in the above statement (b) 
concerning the meaning of the no-interaction theorem,
the invariance of null cones is a more elegant assumption
in view of relativistic invariance.

The plan of this paper is as follows.
In \S 2, we introduce the notation and present the main theorem.
In \S 3, we give a proof of this theorem.
Section 4 is  dovoted to discussion.
In the Appendix, we prove lemmas that are used in \S 3.

\section{Theorem}
The stage of the theory is four-dimensional Minkowski 
affine space $(M,V)$,
where the base set $M$ is the set of all events, and 
the four-dimensional vector space $V$ is the set of all parallel 
transformations of $M$.
The metric of $V$ is 
of type $(+,-,-,-)$ and 
denoted by $ \langle | \rangle $. 
The ordered pair $(m,X)$ of a positive number $m$ and 
a smooth timelike worldline \cite{Yoneda93}
$X$ is called ``the particle".
In this paper, the particle has positive constant rest mass $m$ 
and is structureless.
The worldline $X$ is parametrized by its proper time $\tau$.
Thus $\dot{X}(\tau)$ is the proper velocity, 
which is a future time-like unit, 
$\ddot{X}$ is the proper acceleration, 
$m\dot{X}(\tau)$ is the linear momentum,
and $m\overrightarrow{OX(\tau)} \wedge\dot{X}(\tau)$ 
is the angular momentum 
with respect to the origin $O \in M$,
where $\wedge$ is the ordinary wedge product.
(The arrow denotes a vector.)
The set of particles $(m_i,X_i)$ $(i=1,2,\cdots,N)$
is called the system of $N$ particles 
if they do not collide with each other,
i.e., $X_i \cap X_j = \emptyset$ for $i \neq j$. 

The past null cone with apex $x\in M$ is defined as
\[
C(x)=\{ y\in M:\overrightarrow{yx} \mbox{ is future null }\}.
\]
For a system of $N$ particles, we define 
the linear and angular momenta on the null cones $C(x)$ as
the instantaneous sums over the respective quantities of 
individual particles:
\begin{eqnarray*}
\vec{P}(x)&=&\sum_{i=1,\cdots,N} m_i \dot{X}_i(\tau_i), \\
\mat{A}(x)&=&\sum_{i=1,\cdots,N} m_i \overrightarrow{OX_i(\tau_i)} 
\wedge \dot{X}_i(\tau_i).
\end{eqnarray*}
Here $X_i(\tau_i) \in C(x)$.
(The boldface denotes a rank-two tensor.)
Now we assume that the worldlines intersect 
$C(x)$.
(There is the possibility that 
worldlines do not intersect $C(x)$,
but such a case is not realistic. \cite{Yoneda93})
The no-interaction theorem in this paper is the following.
\\
{\bf Theorem} \quad
{\it For a system of four particles,
if $\vec{P}(x)$ and $\mat{A}(x)$ are independent of $x$,
the worldline of every particle is straight. }

In this paper, we assume an invariant momentum on null cones
instead of on space-like planes, 
which we assumed in the previous paper.

\section{Proof of the theorem}
We can regard $\tau_i$, defined by $X_i(\tau_i)\in C(x)$,
as a scalar field $M \to {\bf R}$. 
From the null condition
$0= \langle 
\overrightarrow{X_i(\tau_i)x}|
\overrightarrow{X_i(\tau_i)x}
\rangle$,
we obtain the gradient of 
$\tau_i$:
\begin{equation}
{\rm grad}\ \tau_i=
\overrightarrow{X_i(\tau_i)x} \; / \;
\langle 
\overrightarrow{X_i(\tau_i)x}|
\dot{X}_i(\tau_i)
\rangle .
\label{gradtau}
\end{equation}

We denote the invariance of the linear momentum 
in tensor form as follows.
We perturb the apex $x$ in the direction $\vec{v} \in V$.
We have
\begin{eqnarray*}
0&=&
\partial \vec{P}(x+\epsilon \vec{v}) /\partial \epsilon
=
\Sigma m_i \langle {\rm grad} \tau_i|\vec{v} \rangle \ddot{X}_i
=
\Sigma m_i
{\langle \overrightarrow{X_ix}|\vec{v}\rangle \over \langle 
\overrightarrow{X_ix}|\dot{X}_i\rangle}\ddot{X}_i
=
\Sigma 
\langle \overrightarrow{X_ix}|\vec{v} \rangle \vec{w}_i,
\end{eqnarray*}
where $\vec{w}_i:={m_i \ddot{X}_i /\langle \overrightarrow{X_ix}|
\dot{X}_i\rangle}$.
(The index $i=1,2,3,4$ of the summation $\Sigma$ 
and the argument $\tau_i$ of the functions 
$X_i, \dot{X}_i$ and $\ddot{X}_i$
are omitted from this point.)
Due to the arbitrariness of $\vec{v}$, 
we see that the tensor
\begin{equation}
0=
\mat{Q}
:=
\Sigma 
\overrightarrow{X_ix} \otimes \vec{w}_i
\end{equation}
must vanish,
where $\otimes$ is the ordinary tensor product.

Similarly, we denote the invariance of the angular momentum 
in tensor form as
\begin{eqnarray*}
0&=&
\partial \mat{A}(x+\epsilon \vec{v}) /\partial \epsilon
=
\Sigma 
m_i\langle {\rm grad}\tau_i|\vec{v}\rangle \dot{X}_i \wedge  
\dot{X}_i
+
m_i\overrightarrow{OX_i} \wedge \langle {\rm grad}\tau_i|\vec{v}\rangle 
\ddot{X}_i
\\&=&
\Sigma 
\overrightarrow{OX_i} \wedge \langle \overrightarrow{X_ix}|
\vec{v}\rangle \vec{w}_i.
\end{eqnarray*}
Due to the arbitrariness of $\vec{v}$, we thus see that
the following tensor must vanish:
\begin{equation}
0=\mat{B}:=
\Sigma
\overrightarrow{X_ix} \otimes \overrightarrow{OX_i} \wedge \vec{w}_i.
\end{equation}

We will prove $\ddot{X}_1(\tau_1)=0$.
The number $i=1$ and proper time $\tau_1$ 
do not have special meaning.
Thus the proof of $\ddot{X}_1(\tau_1)=0$ is sufficient to obtain
the conclusion of the theorem.

We may choose $x$ so that
$\overrightarrow{X_1x}$
is parallel with neither 
$\overrightarrow{X_2x}$,
$\overrightarrow{X_3x}$
nor
$\overrightarrow{X_4x}$.
(See Lemma \ref{nonpara}.)
It is useful to consider the independence of 
the null vectors
$\overrightarrow{X_1x}$, $\overrightarrow{X_2x}$,
$\overrightarrow{X_3x}$ and
$\overrightarrow{X_4x}$ 
by considering four cases:
(a) $\dim[\overrightarrow{X_2x}$,$\overrightarrow
{X_3x} $,$\overrightarrow{X_4x}]=1$;
(b) $\dim[\overrightarrow{X_2x}$,$\overrightarrow
{X_3x}$,$\overrightarrow{X_4x}]=2$;
(c) $\dim[\overrightarrow{X_1x}$,$\overrightarrow
{X_2x}$,$\overrightarrow{X_3x}$,$\overrightarrow{X_4x}]=4$
(so $\dim[\overrightarrow{X_2x}$,$\overrightarrow
{X_3x}$,$\overrightarrow{X_4x}]=3$);
(d) 
$\dim[\overrightarrow{X_2x}$,$\overrightarrow
{X_3x}$,$\overrightarrow{X_4x}]=3$ and
$\dim[\overrightarrow{X_1x}$,$\overrightarrow
{X_2x}$,$\overrightarrow{X_3x}$,$\overrightarrow{X_4x}]=3$.

For the cases (a), (b) and (c),
we see that
there exists $\vec{r}\in V$ such that
\begin{equation}
\langle\overrightarrow{X_1x}|\vec{r}\rangle\neq 0
\ \mbox{ and }\ 
\langle\overrightarrow{X_2x}|\vec{r}\rangle=
\langle\overrightarrow{X_3x}|\vec{r}\rangle=
\langle\overrightarrow{X_4x}|\vec{r}\rangle=0.
\label{ar}
\end{equation}
Thus in these cases,  
using $\mat{Q}(\vec{r},\cdot)=0$,
we have $\vec{w}_1=0$, and thus $\ddot{X}_1(\tau_1)=0$.

For the case (a),
the three vectors $\overrightarrow{X_2x}$, $\overrightarrow{X_3x}$
and $\overrightarrow{X_4x}$
are parallel to each other,
and not parallel to $\overrightarrow{X_1x}$.
Thus $\vec{r}=\overrightarrow{X_2x}$ satisfies (\ref{ar}).

For the case (b),
some pair of $\overrightarrow{X_2x}$, $\overrightarrow{X_3x}$
and $\overrightarrow{X_4x}$
are parallel to each other,
while the remaining one is 
not parallel to the others,
because 
three non-parallel null vectors are linearly independent
(Lemma \ref{3null}).
We assume, for example, 
$\overrightarrow{X_2x} \not\parallel \overrightarrow{X_3x} 
 \parallel \overrightarrow{X_4x}$.
Then let us set
\begin{eqnarray*}
\vec{r}&:=&
-\langle \overrightarrow{X_2x}|\overrightarrow{X_3x} \rangle 
 \overrightarrow{X_1x}
+\langle \overrightarrow{X_1x}|\overrightarrow{X_3x} \rangle 
 \overrightarrow{X_2x}
+\langle \overrightarrow{X_1x}|\overrightarrow{X_2x} \rangle 
 \overrightarrow{X_3x}.
\end{eqnarray*}
We then easily see this $\vec{r}$ satisfies (\ref{ar}).

For the case (c),
let us consider a nonzero vector 
$\vec{r}\in [\overrightarrow{X_2x},
\overrightarrow{X_3x},\overrightarrow{X_4x}]^\perp$
(orthogonal space).
We remark that
$V=[\overrightarrow{X_1x},\overrightarrow{X_2x},\overrightarrow{X_3x},
\overrightarrow{X_4x}]$.
If we assume, on the contrary, that 
$\langle \vec{r} | \overrightarrow{X_1x}\rangle=0$,
the non-zero vector $\vec{r}$ is perpendicular to all vectors of $V$,
which contradicts the non-degeneracy of the metric.
Therefore we see that $\vec{r}$ satisfies (\ref{ar}).

The remaining case is (d).
The proof for this case is more complicated than the others.
Since 
$\dim[\overrightarrow{X_2x},\overrightarrow{X_3x},
 \overrightarrow{X_4x}]=3$,
we see that none of 
$\overrightarrow{X_2x}$, $\overrightarrow{X_3x}$ and 
$\overrightarrow{X_4x}$ 
are parallel to each other.
Thus none of 
$\overrightarrow{X_1x}$, $\overrightarrow{X_2x}$, 
$\overrightarrow{X_3x}$
and $\overrightarrow{X_4x}$ 
are parallel to each other.
Since
$\dim[\overrightarrow{X_1x},\overrightarrow{X_2x},
\overrightarrow{X_3x},\overrightarrow{X_4x}]=3$,
there exist nonzero real $a_i$ $(i=1,2,3,4)$ such that
$\Sigma a_i \overrightarrow{X_ix}=0$.
Note that if we have $\Sigma b_i\overrightarrow{X_ix}=0$,
then there exists $b \in {\bf R}$ such that
$b_i=a_ib$, because of the linear independence of 
$\overrightarrow{X_2x}$, $\overrightarrow{X_3x}$ 
and $\overrightarrow{X_4x}$.
(Lemma \ref{abratio})

{}From $\mat{Q}=0$, 
there exists $\vec{w}\in V$ 
such that $\vec{w}_i=a_i \vec{w}$.
(Use Lemma \ref{abratio} for every component of $\vec{w}_i$.)
By substitution of $\vec{w}_i=a_i \vec{w}$ into $\mat{B}=0$, 
we have
\begin{eqnarray*}
0&=&
\mat{B}=
\Sigma
\overrightarrow{X_ix} \otimes  
\overrightarrow{OX_i} \wedge a_i \vec{w}.
\end{eqnarray*}
Thus we see that
there exists ${\sf C} \in  V\wedge V$ such that
$\overrightarrow{OX_i} \wedge a_i \vec{w}=a_i{\sf C}$ $(i=1,2,3,4)$.
(Use Lemma \ref{abratio} for every component of 
$\overrightarrow{OX_i} \wedge a_i \vec{w}$.)
Therefore we see that
$\overrightarrow{OX_i} \wedge \vec{w}$ is independent of $i$.
Thus we have
$\overrightarrow{X_iX_j}\wedge\vec{w}=0$ $(i,j=1,2,3,4)$.
This implies that
$\overrightarrow{X_iX_j}\parallel \vec{w}$ $(i,j=1,2,3,4)$.
Noting that
$\overrightarrow{X_1X_2}$ is not parallel to 
$\overrightarrow{X_1X_3}$
(If they are parallel, 
$\overrightarrow{X_1x}$, $\overrightarrow{X_2x}$ and 
$\overrightarrow{X_3x}$ are dependent, 
which contradicts Lemma \ref{3null}.),
we get 
$\vec{w}=0$.
Thus we have $\vec{w}_1=0$,
so that $\ddot{X}_1(\tau_1)=0$.
This proves the theorem.

\section{Discussion}
We succeeded in proving the no-interaction theorem with 
the assumption of the invariance of the momentum on null cones.
Because of differences in the assumptions,
we do not give another proof of 
the no-interaction theorems of the papers cited.
However, we belive that our theorems (in this and previous papers) 
follow the spirit of the Currie no-interaction theorem.
The reason that the proof can be carried out in the same way
as that for the previous paper's theorem is that
there are four particles.
The following two questions are still not answered
and are being studied:
(1) Is the theorem satisfied in the case of more than five particles?
(2) Is the assumption for the angular momentum necessary?


\appendix
\section{}
\begin{lemma}\label{nonpara}
There exists $x_0\in M$ such that
$\overrightarrow{X_1(\tau_1)x_0}$
is not parallel with 
$\overrightarrow{X_2(\tau_2)x_0}$,
$\overrightarrow{X_3(\tau_3)x_0}$
nor
$\overrightarrow{X_4(\tau_4)x_0}$,
where
$X_1(\tau_1), X_2(\tau_2), X_3(\tau_3), X_4(\tau_4) \in C(x_0)$.
\end{lemma}

{\it proof} \quad
Let us choose 
$\tau'_2,\tau'_3,\tau'_4$ so that
$X_2(\tau'_2),X_3(\tau'_3),X_4(\tau'_4)\in C(X_1(\tau_1))$.
We can choose the future null vector 
$\vec{p}$,
except for the future null directions
$\overrightarrow{X_2(\tau'_2)X_1(\tau_1)}$,
$\overrightarrow{X_3(\tau'_3)X_1(\tau_1)}$
and 
$\overrightarrow{X_4(\tau'_4)X_1(\tau_1)}$.
Also, we set $x_0=X_1(\tau_1)+\vec{p}$.
Then $X_1(\tau_1)\in C(x_0)$.
Next, choose $\tau_2,\tau_3,\tau_4$ such that
$X_2(\tau_2),X_3(\tau_3),X_4(\tau_4)\in C(x_0)$.
If we assumed 
$\overrightarrow{X_1(\tau_1)x_0} 
\parallel
\overrightarrow{X_2(\tau_2)x_0}$, by contrast,
we see
$\overrightarrow{X_2(\tau_2)X_1(\tau_1)}
\parallel 
\overrightarrow{X_1(\tau_1)x_0}
=\vec{p}$,
which contradicts the way in which $\vec{p}$ was selected.
Hence we conclude
$\overrightarrow{X_1(\tau_1)x_0}
\not\parallel
\overrightarrow{X_2(\tau_2)x_0}$.
Similarly we see
$\overrightarrow{X_1(\tau_1)x_0}
\not\parallel
\overrightarrow{X_3(\tau_3)x_0}$,
$\overrightarrow{X_1(\tau_1)x_0}
\not\parallel
\overrightarrow{X_4(\tau_4)x_0}$.
$\Box$

\begin{lemma}\label{3null}
Three null vectors that are not parallel to each other
are linearly independent. 
\end{lemma}

{\it proof} \quad
For three non-parallel null vectors 
$\vec{p}_1$, $\vec{p}_2$ and $\vec{p}_3$,
we assume on the contrary that
$\vec{p}_1=a\vec{p}_2+b\vec{p}_3$ for some nonzero $a,b \in {\bf R}$.
Then we would see 
$\langle \vec{p}_1|\vec{p}_1\rangle 
=
2ab\langle \vec{p}_2|\vec{p}_3\rangle 
\neq 0$, which is a contradiction.
Thus we see that the vectors
$\vec{p}_1$,
$\vec{p}_2$ and 
$\vec{p}_3 $
are linearly independent. $\Box$

\begin{lemma}\label{abratio}
For a
vector $\vec{p}_1$
and three independent vectors
$\vec{p}_2$, $\vec{p}_3$ and $\vec{p}_4$,
we assume that 
there exists a nonzero real $a_i$ $(i=1,2,3,4)$ such that
$\Sigma a_i \vec{p}_i=0$.
If we have
$\Sigma b_i \vec{p}_i=0$,
then 
there exists $b\in {\bf R}$ such that
$b_i=a_i b$ $(i=1,2,3,4)$.
\end{lemma}

{\it proof} \quad
We may assume $b_1\neq 0$,
because if $b_1=0$, we see that
$b_2=b_3=b_4=0$ by the independence of 
$\vec{p}_2$, $\vec{p}_3$ and $\vec{p}_4$.
Then we can set $b=0$.
From $\Sigma a_i\vec{p}_i=0$ and $\Sigma b_i\vec{p}_i=0$, we have
\begin{eqnarray*}
\vec{p}_1&=&
-{a_2\over a_1}\vec{p}_2
-{a_3\over a_1}\vec{p}_3
-{a_4\over a_1}\vec{p}_4
=
-{b_2\over b_1}\vec{p}_2
-{b_3\over b_1}\vec{p}_3
-{b_4\over b_1}\vec{p}_4,
\\
0&=&
 \left({a_2\over a_1}-{b_2\over b_1}\right)\vec{p}_2
+\left({a_3\over a_1}-{b_3\over b_1}\right)\vec{p}_3
+\left({a_4\over a_1}-{b_4\over b_1}\right)\vec{p}_4.
\end{eqnarray*}
Since the vectors $\vec{p}_2$, $\vec{p}_3$ and $\vec{p}_4$ 
are linearly independent,
we get $a_2/a_1=b_2/b_1$, $a_3/a_1=b_3/b_1$ and $a_4/a_1=b_4/b_1$.
Hence we obtain
$b_1/a_1=b_2/a_2=b_3/a_3=b_4/a_4$.
Therefore we can set $b=b_1/a_1$.
$\Box$

\end{document}